%% file: paper.tex
\documentclass[sigconf]{acmart}
\def\BibTeX{{\rm B\kern-.05em{\sc i\kern-.025em b}\kern-.08emT\kern-.1667em\lower.7ex\hbox{E}\kern-.125emX}}
    
\usepackage{nicefrac}
\usepackage{siunitx}
\usepackage{array,framed}
\usepackage{booktabs}
\usepackage{
  color,
  float,
  epsfig,
  wrapfig,
  graphics,
  graphicx,
  subcaption
}

\usepackage{textcomp,amssymb}
\usepackage{setspace}
\usepackage{latexsym,fancyhdr,url}
\usepackage{enumerate}
\usepackage{algorithm2e}
\usepackage{algpseudocode}
\usepackage{graphics}
\usepackage{xparse} 
\usepackage{xspace}
\usepackage{multirow}
\usepackage{csvsimple}
\usepackage{soul}
\usepackage{balance}

\usepackage{subcaption}
\usepackage{subcaption}
\usepackage{graphicx}
\usepackage{svg}

\usepackage{
  tikz,
  pgfplots,
  pgfplotstable
}
\usepackage{hyperref}

\usetikzlibrary{
  shapes.geometric,
  arrows,
  external,
  pgfplots.groupplots,
  matrix
}

\pgfplotsset{compat=1.9}

\usepackage{mathtools,}

\DeclareMathAlphabet{\mathcal}{OMS}{cmsy}{m}{n}

\DeclareGraphicsExtensions{%
    .png,.PNG,%
    .pdf,.PDF,%
    .jpg,.mps,.jpeg,.jbig2,.jb2,.JPG,.JPEG,.JBIG2,.JB2}

\usepackage{refcount} 

\newcommand{\commercialfn}{%
  \footnote{\label{fn:commercial}In accordance with vendor licensing terms, commercial product identifiers have been redacted and are presented in anonymized form.}%
}

\newcommand{\commercialmark}{%
  \footnotemark[\getrefnumber{fn:commercial}]%
}

\input{defs}
\begin{document}
\fancyhead{}
\def\thetitle{StealthCup: Realistic, Multi-Stage, Evasion-Focused CTF for Benchmarking IDS}
\title{\thetitle}

\author{Manuel Kern, Dominik Steffan, Felix Schuster, Florian Skopik, Max Landauer, David Allison}
\affiliation{\small{Austrian Insititute of Technology}}
\email{firstname.lastname@ait.ac.at}

\author{Simon Freudenthaler}
\affiliation{\small{FH Hagenberg}}
\email{freudenthaler.simon@gmail.com}

\author{Edgar Weippl}
\affiliation{\small{University of Vienna}}
\email{edgar.weippl@univie.ac.at}


\date{}

\begin{abstract}
Intrusion Detection Systems (IDS) are critical to defending enterprise and industrial control environments, yet evaluating their effectiveness under realistic conditions remains an open challenge. 
Existing benchmarks rely on synthetic datasets (e.g., NSL-KDD, CICIDS2017) or scripted replay frameworks, which fail to capture adaptive adversary behavior. 
Even MITRE ATT\&CK Evaluations, while influential, are host-centric and assume malware-driven compromise, thereby under-representing stealthy, multi-stage intrusions across IT and OT domains. 
We present \emph{StealthCup}, a novel evaluation methodology that operationalizes IDS benchmarking as an evasion-focused Capture-the-Flag competition. 
Professional penetration testers engaged in multi-stage attack chains on a realistic IT/OT testbed, with scoring penalizing IDS detections. 
The event generated structured attacker writeups, validated detections, and PCAPs, host logs, and alerts.
Our results reveal that out of 32 exercised attack techniques, 11 were not detected by any IDS configuration. 
Open-source systems (Wazuh, Suricata) produced high false-positive rates ($>$90\%), while commercial tools generated fewer false positives but also missed more attacks. 
Comparison with the Volt Typhoon APT advisory confirmed strong realism: all 28 applicable techniques were exercised, 19 appeared in writeups, and 9 in forensic traces. 
These findings demonstrate that StealthCup elicits attacker behavior closely aligned with state-sponsored TTPs, while exposing blind spots across both open-source and commercial IDS. 
The resulting datasets and methodology provide a reproducible foundation for future stealth-focused IDS evaluation.
\end{abstract}
\maketitle


\section{Introduction}
Intrusion Detection Systems (IDS) remain a cornerstone of enterprise and critical infrastructure defense. 
Yet despite advances in machine learning and analytics, evaluating IDS performance under realistic conditions is still an open problem. 
Most academic work continues to benchmark IDS on synthetic datasets such as NSL-KDD~\cite{LDhanabal2015ASO} or CICIDS2017~\cite{cicds2017}. 
While useful for regression testing, these datasets are artificially generated and diverge statistically from real traffic distributions, limiting their value as indicators of real-world detection effectiveness~\cite{11068309,idssurvey25,sharafaldin2018toward}. 
Replay frameworks and testbeds~\cite{mitreeval,9262078,caldera} offer controlled experimentation but typically rely on scripted exploits, failing to capture adaptive adversary behavior. 
The MITRE ATT\&CK Evaluations~\cite{mitreeval} have become a widely recognized industry benchmark for endpoint detection products, but they are fundamentally host-centric, rely on dated threat intelligence reports, and assume initial compromise via commodity malware—conditions under which EDR solutions naturally excel while network- or anomaly-focused IDS are underrepresented.
This enables vendors to tune products to well-documented attack steps, highlighting coverage of known techniques while under-representing stealthy, multi-phase operations observed in modern campaigns.

In practice, advanced adversaries exploit subtle misconfigurations, credential hygiene issues, and lateral movement chains spanning IT and OT domains \cite{mandiant_mtrends2025,volttyphon,verizon_dbir2025}. 
Such behavior is difficult to reproduce in scripted or malware-centric evaluations, but it is crucial for understanding IDS blind spots in realistic deployments. 
This motivates the need for new evaluation methodologies that (i) capture human attacker ingenuity and evasion, (ii) reflect complex IT/OT infrastructures, and (iii) provide reproducible benchmarks with transparent results.

Based on these challenges, we derive the following research questions that guide our study:
\label{sec:RQs}

\textbf{RQ1:} To what extent does StealthCup generate attacker behavior and datasets that resemble those observed in real-world APT reports, and are thus suitable for IDS evaluation?

\textbf{RQ2:} To what extent can the approach capture the impact of IDS configuration choices (e.g., open-source vs.\ commercial, default vs.\ tuned) on detection effectiveness against stealthy attacker behavior?


\smallskip
\noindent\textbf{Contributions.}  
Our main contributions are as follows:
\begin{enumerate}
  \item \textbf{The StealthCup framework:} an innovative evaluation approach combining evasion-oriented, multi-stage attack challenges in a structured competition format.
  \item \textbf{Validated IT/OT testbed:} implementation of a realistic, vulnerable, and expert-validated IT/OT infrastructure as Infrastructure-as-Code, including documented multi-stage attacks and IDS rule sets.
  \item \textbf{Comparative IDS evaluation:} systematic analysis of open-source (Wazuh, Suricata) and commercial (Vendor A EDR, Vendor B NIDS)\commercialfn{} IDS solutions under realistic stealth scenarios, highlighting strengths, blind spots, and the effect of tuning.
  \item \textbf{Open dataset release:} public release of alerts, event logs, PCAPs, and attacker writeups to facilitate independent verification and future research.
\end{enumerate}

By combining human-driven evasion with reproducible infrastructure and transparent scoring, StealthCup complements existing evaluation efforts (datasets, replay frameworks, and MITRE ATT\&CK Evaluations) with an adversarial perspective that more closely reflects how IDS solutions perform against stealthy, adaptive attackers.

In its inaugural run, StealthCup brought together 52 participants across 12 international teams to attack a realistic IT/OT infrastructure. 
The competition produced 14 structured attacker writeups and PCAPs, alerts, and host logs. 
Analysis showed that 11 of 32 attack techniques evaded all IDS configurations, open-source systems exhibited false-positive rates above 90\%, and commercial tools missed critical techniques despite producing fewer false positives. 
By aligning 28 exercised techniques with the Volt Typhoon APT advisory, StealthCup confirmed that participant behavior closely mirrors state-sponsored tradecraft, underscoring both the realism of the approach and its value for benchmarking IDS under stealth-focused adversaries.



\section{Related Work}
\label{sec:relatedwork}

\paragraph{Benchmarking IDS with datasets.} 
Evaluation of intrusion detection systems has traditionally relied on standardized datasets. 
Early benchmarks such as DARPA98~\cite{darpa98} and its successors provided labeled traffic traces for training and evaluation, but were later criticized for unrealistic traffic generation and over-simplified attack scenarios~\cite{idsbenchmeth,ranum2001experiences}. 
More recent datasets, including NSL-KDD~\cite{LDhanabal2015ASO} and CICIDS2017~\cite{cicds2017}, continue to be widely used for machine learning-based IDS benchmarking~\cite{maseer2021benchmarking,landauer2024introducing}. 
While useful for reproducibility and regression testing, such datasets remain artificially curated, lack attacker adaptiveness, and diverge statistically from real-world traffic distributions~\cite{11068309,idssurvey25}. 
Consequently, high accuracy on synthetic traces does not necessarily transfer to operational environments.

\paragraph{Testbeds and evaluation initiatives.} 
Beyond static datasets, DARPA-style testbeds~\cite{landauer2024introducing} aim to create controlled environments for IDS evaluation. 
These allow experiments with live systems but typically rely on scripted exploits and predefined attack scenarios. 
MITRE ATT\&CK Evaluations~\cite{mitreeval} represent a widely recognized industry effort, mapping adversary emulation to ATT\&CK techniques. 
However, they are host-centric, rely on dated threat intelligence reports, and assume initial access via commodity malware. 
As a result, endpoint detection and response (EDR) tools tend to score well, while network-based IDS and anomaly-focused approaches are underrepresented. 
Furthermore, because evaluation steps are publicly documented, vendors can tune their systems in advance, limiting insights into evasive or novel attacker behavior.

\paragraph{Red teaming and capture-the-flag (CTF).} 
Professional red-team exercises provide the closest analogue to real intrusions, as skilled testers emulate advanced adversary tactics to remain undetected. 
However, such engagements are costly, results are rarely disclosed, and reproducibility is limited. 
CTF competitions exist in many formats~\cite{zafar24,defcon2003,davis14,ccdcoe,bock18,ctftime2024,hackasat,pwn2own,kucek20}, including attack--defense and defend-only types, but they generally do not focus on systematically evaluating IDS configurations. 
Academic experiments have explored integrating IDS into competitions, e.g., the UCSB iCTF 2009 where IDS alerts influenced scoring~\cite{ictf-snort}, but IDS played only a secondary role and experimental control was limited. 
More broadly, CTFs such as DEFCON CTF~\cite{defcon2003} or Pwn2Own~\cite{pwn2own} highlight the creativity of adversarial participants, but do not provide structured IDS evaluation. 
Notably, prior work and the MITRE ATT\&CK framework~\cite{mitreeval} itself underscore that penetration testers and advanced persistent threats (APTs) share overlapping tactics, techniques, and procedures (TTPs). 
Both seek stealth, persistence, lateral movement, and privilege escalation, while routinely attempting to evade security controls. 
Thus, CTF-style engagements provide a natural mechanism to elicit attacker ingenuity similar to state-sponsored campaigns.

\paragraph{Positioning StealthCup.} 
StealthCup complements these approaches by combining the creativity of human attackers with the reproducibility of controlled testbeds. 
Unlike synthetic datasets (DARPA98, CICIDS2017), StealthCup generates live attack traces focused explicitly on detection evasion. 
Unlike MITRE ATT\&CK Evaluations, its objectives are not fully disclosed in advance, incentivizing adaptive and stealthy attacker strategies. 
And unlike prior CTFs, StealthCup treats IDS evasion as a first-class objective with structured scoring, reproducible Infrastructure-as-Code deployments, and public release of alerts, logs, and attacker writeups. 
This positioning makes StealthCup a complementary methodology that reveals IDS blind spots specific to configuration and deployment, while ensuring comparability and transparency.




\section{Methodology}
StealthCup integrates Capture-the-Flag (CTF) gamification into intrusion detection benchmarking. The framework is designed to capture realistic attacker behavior, evaluate IDS resilience against stealthy operations, and produce reproducible datasets. 

\subsection{Overview of the StealthCup Framework}
StealthCup provides a generic evaluation framework that can be instantiated in different domains and infrastructures. 
A concrete instantiation is detailed later in the illustrative use case (Sect.~\ref{sec:ill-uc}), whereas this section outlines the core methodology. 
At a high level, the framework consists of four key steps:

\paragraph{Competition Design.}
StealthCup adopts an \emph{attack-only evasion focused CTF} format. Participants are tasked with solving multi-stage attack challenges while avoiding detection. Every triggered alert while solving a challenge incurs a penalty. This shifts the incentive structure: success is measured not by how quickly teams achieve objectives, but by how stealthily they can operate compared to their competitors solving a challenge. 

To make this effective, StealthCup implements a dynamic scoring scheme. Alerts are weighted by severity, with high-severity alerts generating stronger penalties than low-severity ones. Because severity taxonomies differ between IDS solutions (e.g., Wazuh logs many low-severity alerts unrelated to real compromise, while Vendor A EDR\commercialmark{} low-severity alerts are often true indicators of intrusion), we normalize penalties across systems by applying adjustable weights per IDS. This normalization ensures comparability while preserving fairness across heterogeneous detection technologies. 

\paragraph{Infrastructure and Attack Scenarios.}
Narrative \emph{scenarios} with clearly defined objectives form the foundation of StealthCup. 
Rather than providing isolated exploits or disconnected technical tasks, each scenario consists of multiple \emph{objectives} that mimics a realistic intrusion campaign. 
Objectives are defined in terms of attacker goals (e.g., establish persistence in the Active Directory domain, exfiltrate data, manipulate a PLC register), ensuring that every step contributes to a logically consistent multi-stage chain, such as client compromise $\rightarrow$ Active Directory escalation $\rightarrow$ lateral movement $\rightarrow$ PLC manipulation. 
This design discourages “point-and-shoot” exploitation and instead drives participants toward strategic, multi-phase operations that mirror how real adversaries pursue end-to-end objectives.
To preserve realism while remaining feasible, scenarios incorporate realistic misconfigurations and recently disclosed vulnerabilities (\emph{attack vectors}), reflecting what professional penetration testers would encounter in practice. 
This avoids the disincentive of “burning” valuable zero-day exploits, although we do not prevent participants from employing them. 
We argue that requiring zero-day exploits for operating systems or common software privilege escalation would discourage participation, given their high market value (e.g., Zerodium pricing \cite{zerodium}). In contrast, StealthCup explicitly encourages the discovery or use of zero-day techniques targeting IDS components, as these directly contribute to winning the challenge.

The underlying infrastructure is tailored to the scenario context. 
For instance, an energy-sector case combines enterprise IT assets (Windows domains, Linux servers) with representative OT equipment and protocols (e.g., PLC controllers using Modbus) common in critical infrastructure \cite{colonial_pipeline}. 
In StealthCup, the infrastructure mirrors the socio-technical environment relevant to the chosen domain, while detailed vendor and configuration choices are deferred to the illustrative case study in Section~\ref{sec:infra}. 

In the following, we use the we use the term \emph{testbed} to denote the controlled experimental infrastructure.
In StealthCup, a testbed comprises not only the systems and software, but also their specific configuration state and monitoring solutions. 
This includes both secure baseline setups and deliberately injected misconfigurations or recently disclosed vulnerabilities that serve as attack vectors. 
By explicitly modeling configuration and misconfiguration within the testbed, we can design repeatable attack chains, ensure that all participants face identical opportunities for compromise, and systematically study how IDS solutions respond to different phases of the intrusion. 
The testbed is fully automated using Infrastructure-as-Code (Terraform, Ansible), enabling consistent rebuilds, controlled randomization, and comparability across teams and scenarios.

\paragraph{Data Collection and Ground Truth.}
All experiments are instrumented to collect comprehensive traces, including full packet captures (PCAPs), host logs, and IDS alerts. A key challenge is establishing ground truth in the absence of detailed logging. Installing logging agents on participant machines would provide precise traces, but we deliberately avoid this to not discourage participation or bias attacker behavior. Instead, we combine (i) structured attacker writeups, required upon challenge completion, (ii) manual verification of reported steps against collected telemetry.

While attacker-side logging agents would enable additional KPIs such as time-to-detect and enable throughout labelling, we avoided this in the first StealthCup to not discourage participation; this remains a planned extension.

\paragraph{Evaluation Mechanics.}
To prevent discouragement after early mistakes, teams are allowed to reset their infrastructure on demand. Resets introduce trade-offs: information obtained in a previous run may remain useful, so we implement randomization of hostnames, services, and network configurations to mitigate knowledge carryover. 

Prevention mechanisms (e.g., anti-malware blocking of tools such as Mimikatz) are deliberately disabled, ensuring that participants can operate freely. However, all such actions still generate detection alerts, which are penalized in scoring.

Importantly, detection events are made visible to participants in near-real time, enabling them to adapt their strategies dynamically. 
We adopt this design for three main reasons. 
First, near-real-time feedback helps participants use the limited competition time more effectively, as they can adjust their approach rather than unknowingly persisting with unproductive attack paths. 
Second, it encourages an iterative style of engagement, where participants explore alternative tactics when their initial attempts trigger alerts. 
Finally, from a benchmarking perspective, exposing detections introduces adaptive pressure on IDS solutions, allowing us to observe not only how well they detect first attempts but also how resilient they remain once attackers begin to modify their behavior.

This adaptive interaction loop not only mirrors real-world adversaries but also increases the diversity of collected attack traces, since participants are incentivized to refine and vary their tactics.


\subsection{Ethics}
All activities occur within isolated ranges under informed consent. No production assets or customer data are involved; outbound connectivity is restricted. We log only range telemetry; personal data from attackers is not retained beyond participation records. Released artifacts include Infrastructure as Code, attack-artefacts, IDS configs (default/tuned), scoring code, scenario descriptions with ATT\&CK mappings and sanitized PCAP/log bundles. All data was collected under research ethics approval.


\section{Illustrative case study}
\label{sec:ill-uc}
This section describes how we instantiated the StealthCup framework in a realistic IT/OT environment. 
We outline the design rationale, testbed construction, IDS integration and event preparation of the first StealthCup event. 

\subsection{Scenario Selection and Design Rationale}
We chose a hybrid IT/OT setting to capture both enterprise and industrial intrusion vectors. 
IT infrastructures with Active Directory (AD) domains, Windows, and Linux servers are ubiquitous across enterprises, while OT components are highly relevant for critical infrastructures \cite{volttyphon}.  
The only preventive network security function implemented was basic layer~4 firewalling, ensuring segmentation but no active intrusion prevention.
The initial attacker foothold was implemented as a \emph{Kali} (Kali Linux 2024.3.0) in the enterprise zone. 
This choice reflects common compromise entry points (e.g., a contractor laptop, phishing victim, or insider device) and provides participants with a familiar penetration testing toolset~\cite{verizon_dbir2025,mandiant_mtrends2025, volttyphon}.
This ensures accessibility, comparability across teams, and a controlled starting point for all attacks.

The initial foothold was fixed across all teams. 
Access to the OT environment was only possible by leveraging misconfigurations in the IT/OT-DMZ, requiring skilled participants with expertise in lateral movement and cross-domain trust exploitation.

\subsection{Testbed Creation}
To guide the design of attack vectors, we conducted a survey of \emph{state-of-the-art attacks} on AD networks and OT networks \cite{steffan2025ad, freudenthaler2025ot}.

\paragraph{Threat Model and Assumptions}
\label{sec:threat-model}
We model skilled, human red teams pursuing predefined objectives while minimizing detections. The testbed embeds realistic misconfigurations and recently disclosed vulnerabilities; operating-system or core network zero-days are \emph{not required} by design. 

\emph{Core assumptions.}
\begin{itemize}
  \item \textbf{Isolation.} Each team operates in its own range; no cross-team leakage of traffic, credentials, or state.
  \item \textbf{Detect-only.} Active prevention (e.g., IPS drops, malware blocking) is disabled; detections are visible but do not block execution.
  \item \textbf{Initial foothold.} Teams start from a fixed enterprise client with standard reachability but no pre-positioned privileged credentials.
  \item \textbf{Randomization.} Resets generate fresh instances with randomized credentials, while attack paths and objectives remain constant to ensure comparability.
  \item \textbf{Data integrity.} IDS sensors and the SIEM backend are trusted; tampering is prohibited. Full PCAP, host logs, and IDS alerts are recorded continuously and persist across resets.
\end{itemize}

\emph{Design requirements.}
\begin{itemize}
  \item \textbf{Attack vectors.} Techniques reflect current practice (e.g. living-off-the-land (LotL), AD/PKI misconfigurations, credential hygiene issues, OT protocol weaknesses).
  \item \textbf{Attack paths.} Objectives require multi-stage compromise (e.g., AD persistence; PLC manipulation); lateral movement must be (re)performed after each reset.
  \item \textbf{Replayability.} After resets, difficulty and chains remain consistent to permit fair replay and benchmarking.
  \item \textbf{Scoring.} Penalties derive from alert counts and severities per IDS; validations require a short attacker writeup (cf.\ \S\ref{sec:scoring}).
\end{itemize}

\emph{Limitations.}
\begin{itemize}
  \item User behavior is minimal; UEBA or anomaly-only detection approaches are out of scope.
  \item External threat-intel/IOC feeds are not integrated beyond vendor defaults to avoid confounding.
\end{itemize}

\paragraph{Identification of the IT Attack Chain.} 
To ground the IT part of our testbed in realistic enterprise threats, we systematically identified common Active Directory (AD) attack vectors. This process combined empirical data from large-scale AD security audits (PingCastle - a tool that is used by companies to identify Active Directory misconfigurations gave us their metrics for research purposes \cite{pingcastle}) with literature reviews and penetration testing repositories \cite{steffan2025ad}. 
We prioritized attacks enabled by prevalent misconfigurations, such as weak or missing LDAP signing, unconstrained delegation, or legacy authentication protocols. Each candidate technique was mapped to the MITRE ATT\&CK framework and evaluated for both frequency and impact. This resulted in a curated set of techniques such as Kerberoasting \cite{mitrekerberoast}, AS-REP Roasting \cite{mitreasreproast}, SID-History abuse \cite{mitresidhistory}, forced authentication \cite{mitreforcedauthentication}, and password spraying \cite{mitrepasswordspray}, that are not only frequently exploited by attackers in real-world intrusions, but also reproducible in a controlled lab setting. Incorporating those vectors ensured that our AD environment reflects realistic attack surfaces.

Since we did setup an Active Directory in the OT-DMZ, we identified a domain trust vulnerability \cite{mitresidhistory}, and weak access control \cite{mitrelocalaccounts} as potential gateways for attackers to the OT-DMZ. Those misconfigurations serve as a starting point for the OT Attack Chain.

\paragraph{Identification of the OT Attack Chain.} 
To design a realistic OT attack chain, we first surveyed state-of-the-art adversary techniques targeting both IT and OT environments. \cite{freudenthaler2025ot}
The methodology proceeded in three steps: (i) assessing the opportunities provided by the given infrastructure (trust relationships, weak access control), (ii) theoretically developing attack paths based on known adversary techniques, and (iii) embedding typical weaknesses observed in industrial networks as identified in prior literature. 
The resulting chain leveraged native system functions and common misconfigurations rather than unrealistic exploits. 

For example, access into the OT domain was achieved via an IT account valid in both domains, and lateral movement relied on valid credentials. 
Techniques incorporated credential theft, injection attacks against web applications and ICS interfaces and eavesdropping. 
Vulnerabilities such as unencrypted traffic, outdated software, and insecure web services were systematically embedded to reflect the conditions documented in real critical-infrastructure incidents. 
This approach ensured that the OT attack chain remained both feasible for penetration testers and representative of realistic adversary behavior.

Based on these findings, we implemented realistic misconfigurations and recently disclosed vulnerabilities as deliberate attack vectors.

\begin{figure}[htbp]
  \centering
  \includegraphics[width=\columnwidth]{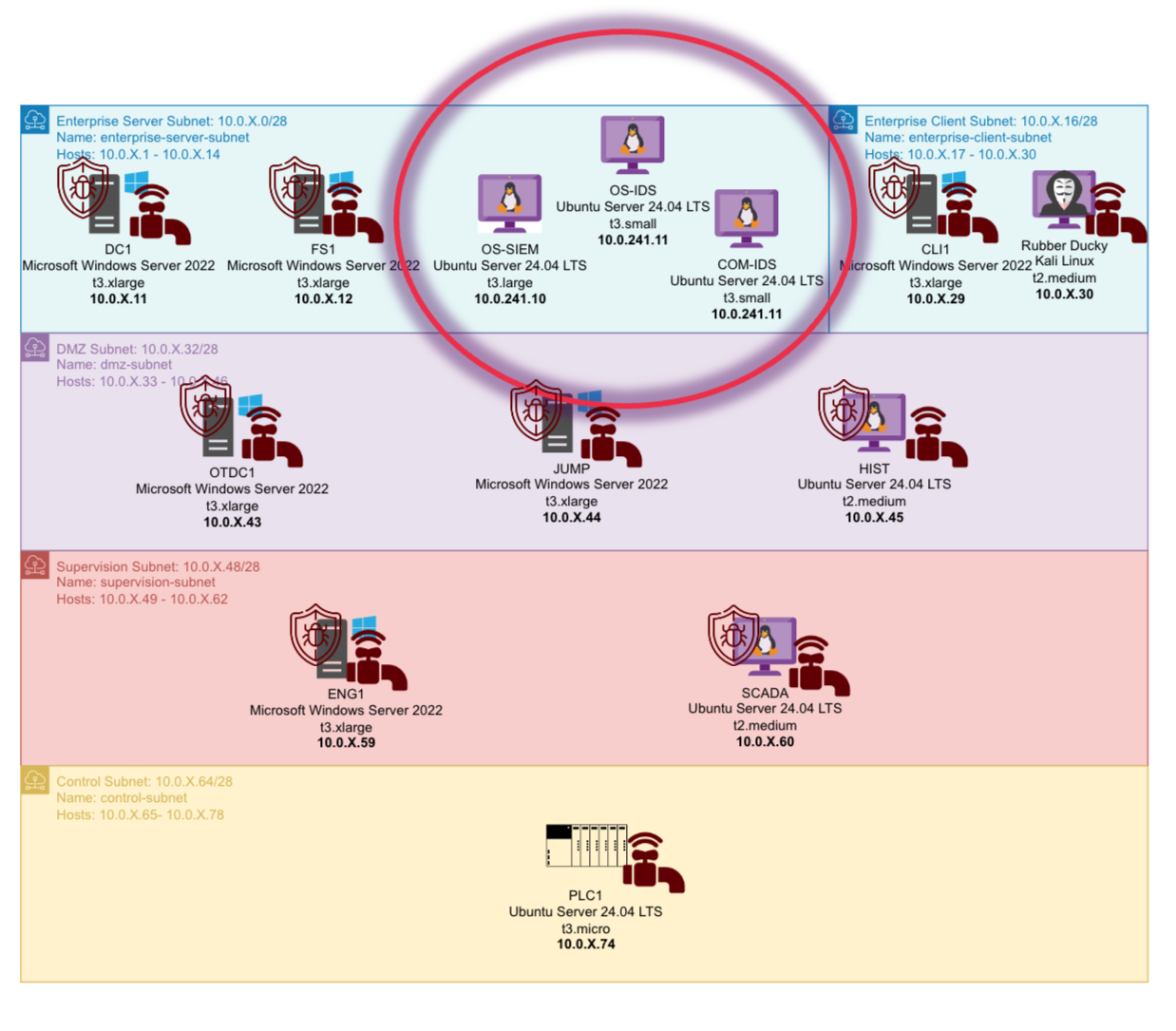}
  \caption{StealthCup multi-level IT/OT infrastructure including NIDS and SIEM in red. The bug symbol indicates installations of HIDS/EDR/XDR, while the tap indicates network-tapping.}
  \label{fig:picture1}
\end{figure}

\paragraph{IT Realism.}
The enterprise IT environment consisted of a classical Active Directory (AD) domain with typical Windows-based components: a domain controller (DC1), a file server (FS1), and a client workstation (CLI1), all running Microsoft Windows Server~2022. 
The AD domain provided central authentication, group policy management, and file sharing services, reflecting a standard enterprise setup. 
This design mirrors common enterprise environments where compromise often begins from a client or contractor system, and attackers subsequently leverage AD misconfigurations or credential abuse to escalate privileges and move laterally.

\paragraph{OT Realism.}
The OT environment was modeled after a segmented industrial control system, including both Windows and Linux-based hosts. 
Core components included an OT domain controller (OTDC1), a jump server (JUMP) for remote access into the OT-DMZ, and a dedicated engineering workstation (ENG1) running \emph{PLCnext Engineer} version~2024.0.4 on Windows Server~2022. 
The engineering workstation provided realistic development and deployment workflows for programming industrial controllers. 
In addition, Linux-based hosts simulated industrial services: a historian server (HIST) built on Grafana~11.5.2 and Telegraf~1.32.3 for telemetry collection, and a SCADA server (SCADA) based on the open-source \emph{ScadaLTS} platform, deployed on Ubuntu Server~24.04 LTS. 
Finally, the core industrial process was represented by a digital twin of a \emph{Phoenix Contact AXC F~2152 PLC}, virtualized by CyberDanube \cite{cyberdanube} with firmware version~2021.6.0 and hardware version~04. 
This PLC supported industrial protocols such as Modbus and included vendor-typical features such as a preconfigured administrative account.

\begin{figure*}[t]
  \centering
  \begin{subfigure}[t]{0.48\textwidth}
    \centering
    \caption{Human-machine interface (HMI) in ScadaLTS to control the virtual PLC.}
    \includegraphics[width=\linewidth]{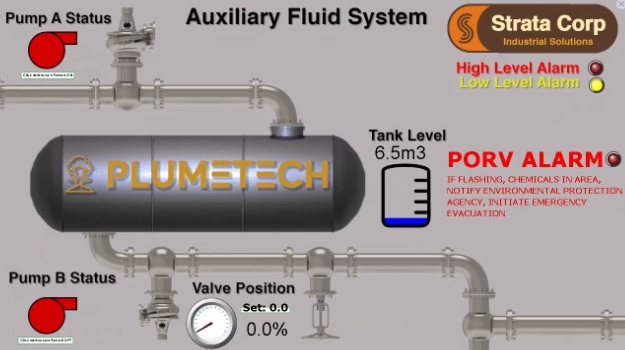}
    \label{fig:plc_hmi}
  \end{subfigure}
  \hfill
  \begin{subfigure}[t]{0.48\textwidth}
    \centering
    \caption{Grafana-based historian logging the emulated I/O states.}
    \includegraphics[width=\linewidth]{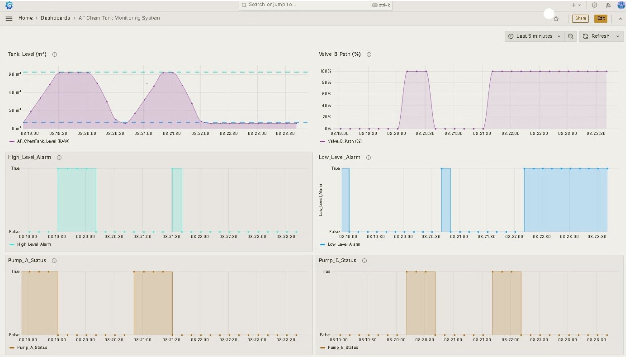}
    \label{fig:plc_historian}
  \end{subfigure}
  \caption{Virtualized PLC controlling a simulated water pump: (a) interaction via ScadaLTS HMI, (b) logging and monitoring via Grafana historian.}
  \label{fig:plc_overview}
\end{figure*}

Furthermore, we programmed the virtualized PLC to simulate the control of a water pump. 
Since the PLC was executed as a QEMU-based virtual machine, no physical I/O was connected, and the sensor and actuator signals were emulated, a technique commonly used in digital twin environments for ICS security research~\cite{stouffer2015purdue,isa95}. 
Pump commands were exposed via Modbus TCP and UDP servers, enabling control through the ScadaLTS HMI~\cite{scadalts} (see Fig.~\ref{fig:plc_hmi}) and recording of pump states in the Grafana-based historian~\cite{grafana} (see Fig.~\ref{fig:plc_historian}).

\subsection{IDS Deployment}
To cover both host- and network-level monitoring, we integrated a mix of commercial and open-source IDS solutions:
\begin{itemize}
    \item \textbf{Host-based IDS (HIDS/EDR):} Vendor~A~EDR\commercialmark{} IDP/XDR (commercial), Wazuh (free open-source). 
    \item \textbf{Network-based IDS (NIDS):} Vendor~B~NIDS\commercialmark{} NIDS (commercial), Suricata (free open-source).
\end{itemize}
IDS sensors were placed across enterprise and OT zones to capture relevant visibility. 

Commercial IDS vendors were approached through direct outreach to industry partners, while open-source solutions were included to ensure reproducibility and comparability. 
Several vendors declined participation for various reasons: Darktrace and ExtraHop emphasized that their anomaly-detection approaches require a minimum deployment size of roughly 500 assets to generate meaningful baselines, making our testbed too small for evaluation. 
Nozomi and SentinelOne did not respond to our inquiries, while Corelight expressed concerns about the uncertainties in the evaluation process. 
IronNet explicitly declined, citing the potential internal costs of participation as a prohibitive factor. 
Ultimately, we integrated two widely used commercial platforms alongside open-source IDS to enable a representative yet feasible evaluation.

Before the competition, we manually tested the predefined attack vectors and authored baseline detection rules for Wazuh and Suricata to ensure that both IDS solutions generated meaningful alerts. 
All attack vectors were successfully detected during these manual tests in the tuned configurations of Wazuh and Suricata.

\subsection{Final Infrastructure and Instrumentation}
\label{sec:infra}

\paragraph{Implementation.}
The full testbed was implemented as Infrastructure-as-Code using Terraform and Ansible. 
This includes enterprise IT components (Windows domains, Linux servers), OT components (emulated PLCs, supervisory servers), and deliberate misconfigurations. 
This approach ensures reproducibility, automated resets, and controlled randomization.

\paragraph{Network Architecture.}
The infrastructure followed the Purdue model, representing state-of-the-art network segmentation in industrial environments~\cite{isa95, stouffer2015purdue}. 
It was divided into four zones: Enterprise IT, OT-DMZ, Supervision, and Control. 
Between each zone, strict layer-4 firewall rules enforced limited connectivity, mirroring typical critical-infrastructure segmentation practices. 
The environment was fully virtualized in AWS, with separate subnets provisioned for each Purdue layer and a consistent addressing scheme to distinguish parallel instances. 
For example, the domain controller in every testbed was assigned the address \texttt{10.0.X.11}, where \texttt{X} denotes the team-specific instance (\texttt{10.0.1.11} for Team~1, \texttt{10.0.2.11} for Team~2, etc.). 
Figure~\ref{fig:picture1} illustrates the overall setup.
The setup was validated in collaboration with four critical-infrastructure operators from the energy and transport sectors to ensure that both IT and OT components reflected real-world deployments.

\paragraph{Validation.}
To ensure realism and reproducibility, all attack techniques were systematically mapped to the corresponding MITRE ATT\&CK tactics, techniques, and sub-techniques. 
For IT-related attacks we used the Enterprise Matrix, while for OT-specific actions we primarily relied on the ICS Matrix. 
Given the close interconnection between IT and OT environments, as also reflected in our testbed, selected Enterprise Matrix techniques were additionally applied to ICS components where relevant.

The theoretically developed attack chain was first reviewed internally with colleagues from professional penetration testing and information security consulting teams to assess both feasibility and operational realism.
Subsequently, three dedicated validation sessions were conducted with OT domain specialists. 
The first session involved experts from T-Systems Austria \cite{tsystems} and CyberDanube \cite{cyberdanube}; the following two involved practitioners from Linz AG \cite{linzag}, Verbund AG \cite{verbundag}, and again CyberDanube \cite{cyberdanube}. 
During these sessions, the proposed attack scenarios and techniques were repeatedly scrutinized with respect to realism and applicability in real-world infrastructures. 
The feedback obtained from these discussions led to refinements of the attack paths.

Finally, all techniques were tested in practice to verify their feasibility within realistic timeframes and under practical preconditions. 
Where necessary, adjustments were made to ensure that each attack step was both technically viable and aligned with adversary behavior observed in operational technology networks. 
This iterative process ensured that the deployed environment and attack chains were not only theoretically grounded but also validated as realistic and relevant to actual critical-infrastructure operations.

\paragraph{Attack Scenario.}
\label{sec:scenarios}
\begin{figure*}[htbp]
  \centering
  \includegraphics[width=\textwidth]{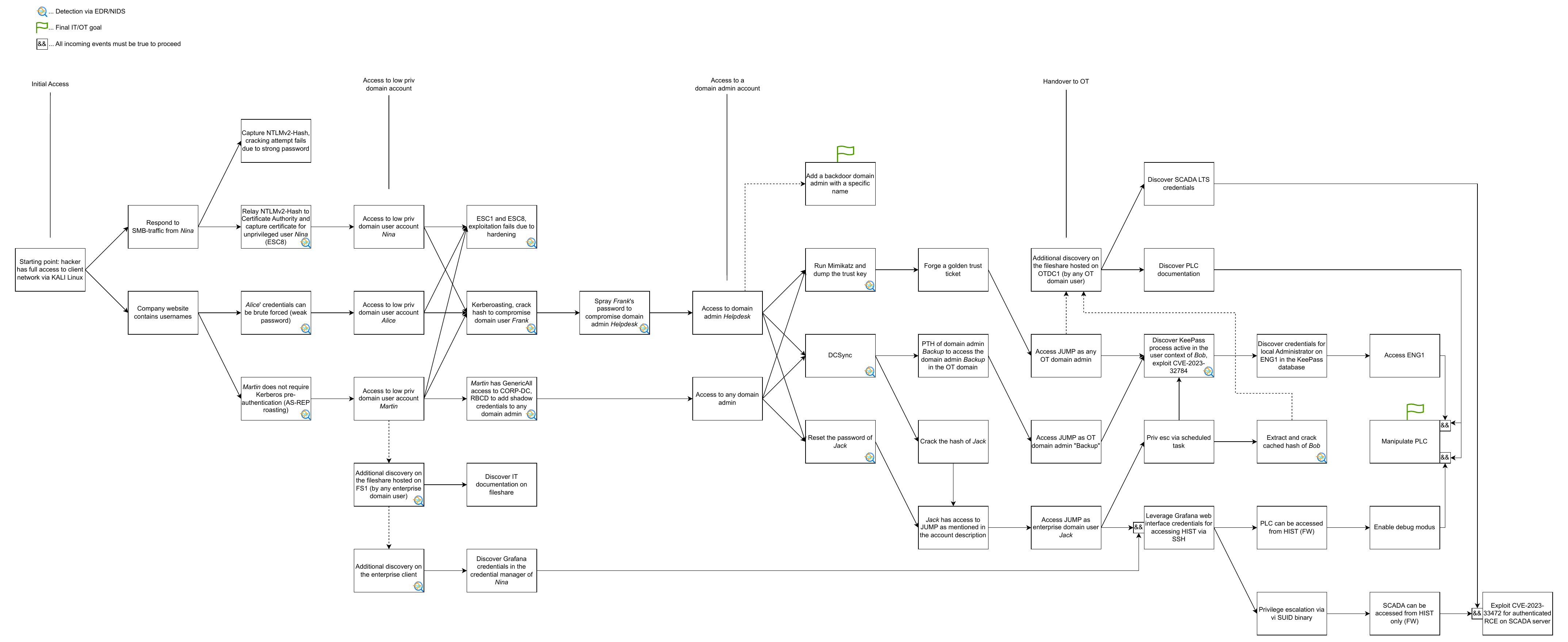}
  \caption{Stealthcup multi staged attack chain that has been implemented including the main objectives and the detections validated in our testrun.}
  \label{fig:picture2}
\end{figure*}

The StealthCup environment supported multiple intrusion paths across IT and OT domains. 
Figure~\ref{fig:picture2} illustrates the implemented multi-stage kill chain with the primary objectives and representative detections. 
The following description highlights one such path, executed by \emph{Team~2}, which successfully completed both objectives.

Initial access was obtained by relaying SMB login requests generated by a scheduled task on the client workstation (CLI1). 
While such credentials are typically harvested through LLMNR/NetBIOS poisoning \cite{mitrellmnrpoisoning} in real-world environments, platform constraints in Amazon Web Services (AWS) required scheduled task-based capture in our setup. 
Using the captured credentials, domain user accounts were enumerated via SID brute-forcing, followed by an AS-REP roasting attack \cite{mitreasreproast} against accounts without Kerberos pre-authentication, which yielded a low-privileged domain account.

Further enumeration of SMB shares on the fileserver exposed sensitive documents and archived emails, including a photograph of the PLC that revealed its default password. 
Privilege escalation was achieved by exploiting an ESC1 misconfiguration \cite{certified_pre_owned} in Active Directory Certificate Services (AD CS). 
This enabled the request of a certificate on behalf of a Domain Administrator, resulting in elevated privileges and persistence through the creation of a new administrator account-fulfilling the IT objective.

The attackers then pivoted to a jump host in the OT-DMZ. 
A misconfigured backup script granted local administrator access, which was exploited to extract a KeePass master password from memory via CVE-2023-32784 \cite{CVE-2023-32784}. 
The recovered database disclosed credentials for lateral movement to the engineering workstation. 
Finally, using the default PLC password identified earlier, Team~2 deployed a modified control project that set the tank-level threshold to zero. 
Uploading this logic disabled safety mechanisms and triggered the simulated release of toxic chemicals, thereby completing the OT objective.

\subsection{Randomization}
During the competition, participants were allowed to reset their infrastructure to recover from mistakes and attempt the objectives again with a fresh score. 
To prevent unfair advantages across resets, all usernames, passwords, and password hashes were randomized, while the overall attack paths and vulnerabilities remained unchanged. 
This ensured that teams could refine their techniques and strategies across successive attempts, but could not simply reuse credentials or artifacts obtained in previous runs. 
For scalability, ten preconfigured environment templates were prepared in advance. 
When a team initiated a reset, its environment was replaced by a freshly randomized instance, and the team was temporarily locked.

\subsection{Competition Design and Scoring}
\label{sec:scoring}
Teams must solve objectives while minimizing IDS detections. 
In the \emph{Enterprise Cup}, the objective is to create a domain administrator account named \emph{plumber} (IT Flag). 
In the \emph{OT Cup}, the objective is to trigger an unsafe state in the PLC such that the process safety valve is activated, simulating a chemical spill (OT Flag). 

\paragraph{Scoring.}
Each team accumulates penalty points based on the number and severity of IDS alerts they trigger. 
All detections are reported in real time on the competition dashboard for all participants. 
Each IDS $k \in \mathcal{K}$ classifies alerts into four severities $s \in \{\text{Critical}, \text{High}, \text{Medium}, \text{Low}\}$. 
Let $a_{t,k,s}$ be the number of alerts of severity $s$ triggered by team $t$, and $w_{k,s}$ the weight assigned to severity $s$ for IDS $k$. 
The penalty is computed as
\[
P_t = \sum_{k \in \mathcal{K}} \sum_{s} w_{k,s} \cdot a_{t,k,s}.
\]
The goal of the competition is to minimize this penalty: the team with the lowest final value wins.

\paragraph{Resets.}
At any time, a team may reset its infrastructure, which restores the environment and the penalty points. 
The first reset is free, but subsequent resets increase a multiplicative factor $\mu_t \geq 1$ applied to the penalty:
\[
\tilde{P}_t = P_t \cdot \mu_t.
\]
Resets also temporarily block access to the infrastructure for 15 minutes, discouraging overuse while still allowing recovery from early mistakes.

\paragraph{Weighting and Validation.}
Weights $w_{k,s}$ were chosen to reflect differing severity scales across IDS solutions. 
For example, Wazuh alerts were weighted $\{50,3,1,0.05\}$ for \{Critical, High, Medium, Low\}, whereas Suricata alerts were weighted $\{500,30,20,10\}$. 
This ensured that severe alerts carried disproportionate penalties and that noisy attacks such as brute-force scanning became non-viable. 

When a team chooses to validate an objective, its current penalty is frozen and must be accompanied by a short attacker write-up describing how IDS detections were bypassed. 
Only validated penalties count for the leaderboard. 
Final rankings are determined by the lowest penalty, with ties broken by (i) fewer hosts accessed, (ii) smaller network footprint, and (iii) shorter time-to-objective.

\subsection{Event Execution}
\paragraph{Event Preparation.}
We advertised the event by presenting StealthCup at security conferences and directly contacting penetration testing companies and security teams. 
The event was designed as a \emph{hybrid competition}, with remote and on-site participation options. 
Rules of the game, the chance of winning four iPhones, as well as subscriptions for hack the box and a narrative storyline (e.g., a fictional company under attack) were published to motivate participants. \cite{github_stealthcup}
Evasion techniques were encouraged, but denial-of-service against IDS components, no tampering with log files, and no exploitation of the gaming backend. 

\paragraph{Participants.}
\begin{table*}[t]
\centering
\scriptsize
\caption{Overview of participating teams. Fields are grouped into professional companies, security consulting firms, and academic researchers. Timestamps shown in UTC.}
\label{tab:participants}
\begin{tabular}{c p{5.2cm} p{3.8cm} c c p{3.2cm} c}
\toprule
\textbf{Team ID} & \textbf{Description of Events} & \textbf{Timestamps (UTC)} & \textbf{Onsite} & \textbf{Remote} & \textbf{Field} & \textbf{Country} \\
\midrule
1  & IT Flag, Reset 1, Writeup 1 & 10:51, 14:28, 15:35 & 8 &  & Security researcher / academic & Austria \\
2  & IT Flag, Reset 1, IT Flag, OT Flag, Writeup 1 & 10:16, 14:11, 15:20, 15:45, 15:55 &  & 4 & Professional company (consulting) & Austria \\
3  & IT Flag, Reset 1, IT Flag, Reset 2, IT Flag, Writeup 1 & 11:42, 12:45, 14:06, 15:28, 16:12, 16:19 &  &  & Professional company (consulting) & UAE \\
4  & IT Flag, Writeup 1, Reset 1, IT Flag, Writeup 2 & 11:27, 12:07, 14:37, 15:48, 16:32 & 4 &  & Security researcher / academic & Austria \\
5  & Credentials OT Hint & 16:06 & 2 &  & Professional company & Austria \\
6  & IT Flag, Writeup 1 & 15:15, 15:38 & 4 &  & Professional company & Austria \\
7  & IT Flag, Reset 1, IT Flag, Writeup 1, Reset 2, IT Flag, Writeup 2 & 13:40, 14:12, 15:15, 15:21, 15:24, 16:12, 16:14 & 4 &  & Professional company (consulting) & Slovenia \\
8  & Reset 1, Reset 2, IT Flag, Writeup 1 & 13:50, 14:58, 16:15, 16:21 & 6 &  & Professional company (consulting) & Austria \\
9  & IT Flag, Writeup 1, Reset 2, IT Flag, Writeup 2 & 10:49, 10:57, 13:50, 14:40, 14:51 & 5 &  & Professional company (consulting) & Austria \\
10 & Writeup 1, Credentials OT Hint, OT Flag, Writeup 2 & 11:53, 16:05, 16:25, 16:29 &  & 6 & Professional company & Netherlands \\
11 & & & 4 & 1 & Security researcher / academic & IRL \\
12 & Writeup 1, Reset 1 & (online w/o timestamp of last edit), 14:57 & 4 &  & Professional company & Austria \\
\bottomrule
\end{tabular}
\end{table*}

The final event included 52 participants, including internationally recognized penetration testers and consulting firms, organized into 12 teams as shown in Table \ref{tab:participants}. 
Teams came from five different countries --- Austria, Ireland, the Netherlands, the United Arab Emirates, and Slovenia --- reflecting both regional and international interest. 
In total, nine teams represented professional security companies, of which five work in security consulting, while three teams were composed of academic researchers. 
Each team received an isolated instance of the IT/OT testbed, consisting of 13 servers and the emulated OT infrastructure. 

In total, 158 virtual machines were deployed in AWS. 
Each team was provided with 13 dedicated hosts, supplemented by two backend machines for orchestration and management.

\section{Data Availability}
\label{sec:dataset}
To ensure reproducibility and foster community use, we release the data and artifacts generated in StealthCup. 
All datasets and code are available at \cite{github_stealthcup}. 
The release includes infrastructure definitions, collected telemetry, and documentation of attack traces, while excluding elements that would pose safety or licensing concerns.

\paragraph{Released Artifacts.}
\begin{itemize}
  \item \textbf{Infrastructure-as-Code.} Full Terraform and Ansible scripts for deploying the IT/OT testbed, including Windows and Linux systems, Active Directory domains, IDS agents/rules, and representative misconfigurations.
  \item \textbf{Datasets.} Complete packet captures (PCAPs), IDS alerts from all deployed solutions, and host logs from Windows and Linux machines.
  \item \textbf{Ground truth.} Structured attacker writeups, submitted upon objective completion, documenting step-wise TTPs. These writeups enable correlation with PCAPs, event logs, and EDR telemetry to identify missed detections.
  \item \textbf{Documentation.} Detailed instructions to reproduce the environments and rerun the experiments.
\end{itemize}

\paragraph{Limitations.}
Commercial software and the PLC digital twin, cannot be open sourced due to vendor licensing; however, equivalent functionality can be replicated with freely available PLCnext software. 
To lower the barrier for experimentation, we selected a low-cost, off-the-shelf PLC controllers.

\section{Evaluation}
\label{sec:RQs}
\subsection{Scoring and Strategies}
We received a total of 14 attacker writeups. 
Eight teams successfully completed the Enterprise (IT) Cup, while only one team, Team~2, reached the OT objective and thereby won the overall competition. 
Team~2 concentrated on achieving full kill chain completion rather than stealth, and consequently ranked last in terms of stealth performance within the IT Cup. 
The main prize (four iPhones for the winning team) was awarded to the overall cup winners. 

The OT challenge required advanced skills and extensive lateral movement. 
From participant feedback and log analysis, we found that only two additional teams had entered the OT environment during regular play. 
At a later stage, we offered selected participants with stronger OT expertise the opportunity to continue exploration by providing credentials to the OT domain, enabling them to escalate further. 

\begin{figure}[bh]
    \centering
    \includegraphics[width=\linewidth]{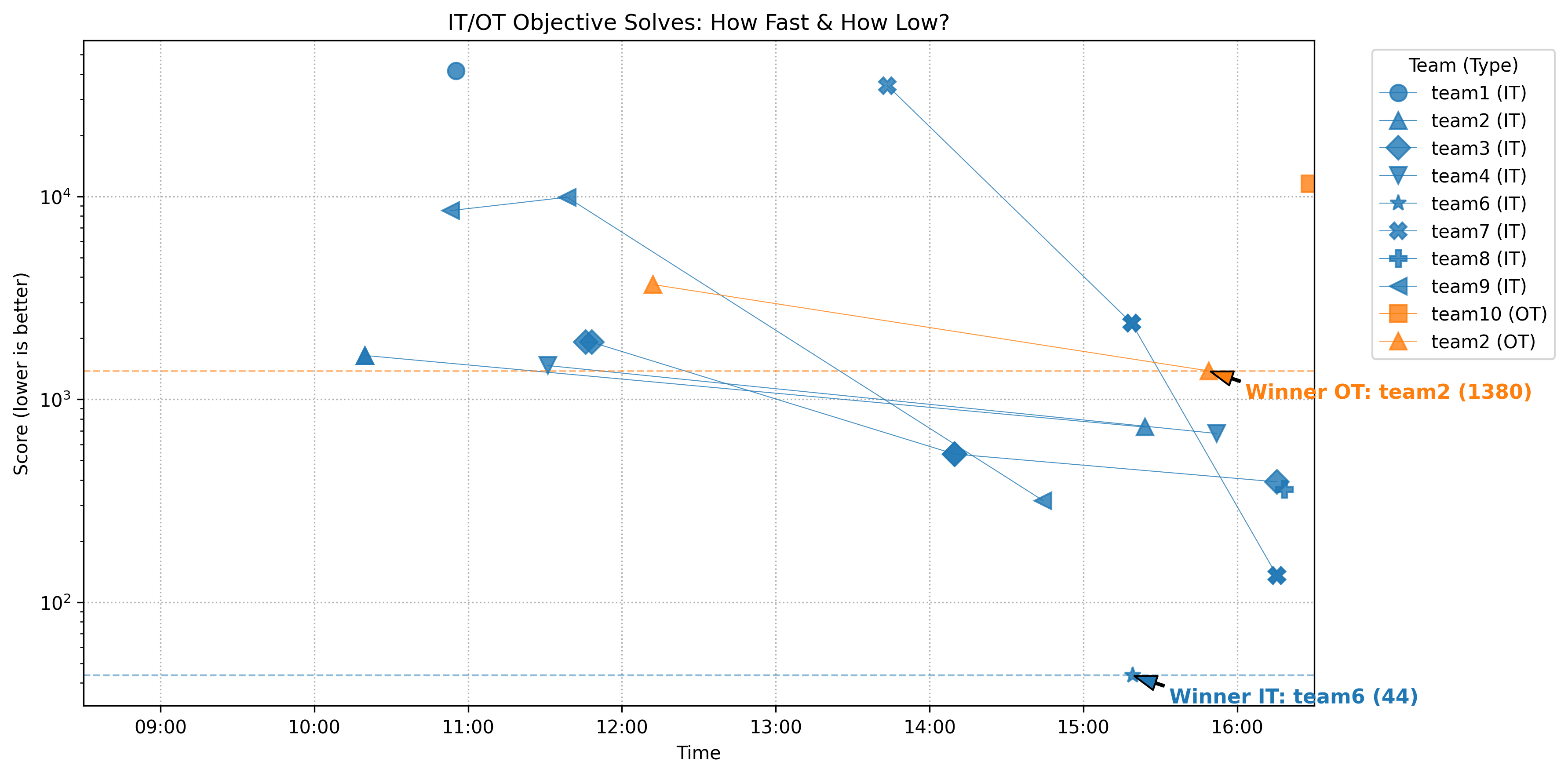}
    \caption{Timeline of the competition, showing objective solves and submitted detection scores (on a logarithmic scale, lower is better).}
    \label{fig:solves}
\end{figure}

These cases are marked as “credentials OT hint” in Table~\ref{tab:participants}. 
Table~\ref{tab:participants} summarizes the participating teams and their progress, while Figure~\ref{fig:solves} illustrates the event timeline, highlighting challenge solves and submitted detection scores (lower is better). 

\subsection{Evaluation of IDS Performance}
\label{sec:evaluation}
After the event, the collected data was analyzed to map attacker actions to IDS detections. 
We reviewed the attacker writeups, attributed each step to the corresponding MITRE ATT\&CK technique, and verified whether alerts were generated by the deployed IDS solutions. 
For each detection, we recorded the severity and whether the technique was identified with precision. 

\begin{figure*}[t]
  \centering
  \begin{subfigure}[b]{0.485\textwidth}
    \centering
    \includegraphics[width=\linewidth]{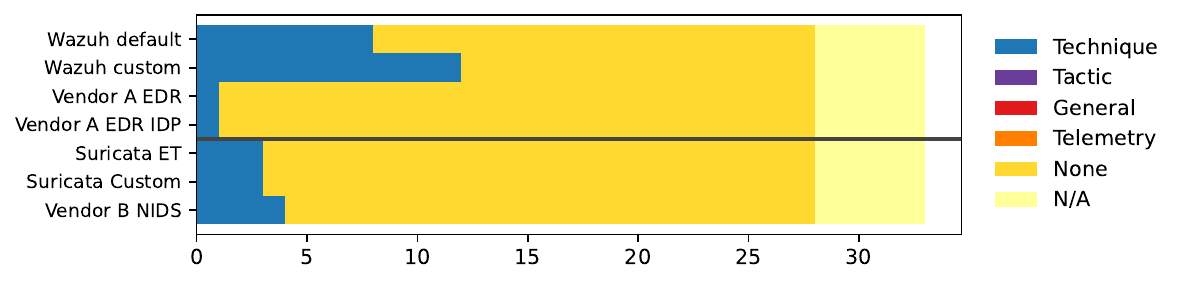}
    \label{fig:mitre_team2}
  \end{subfigure}\hfill
  \begin{subfigure}[t]{0.485\textwidth}
    \centering
    \includegraphics[width=\linewidth]{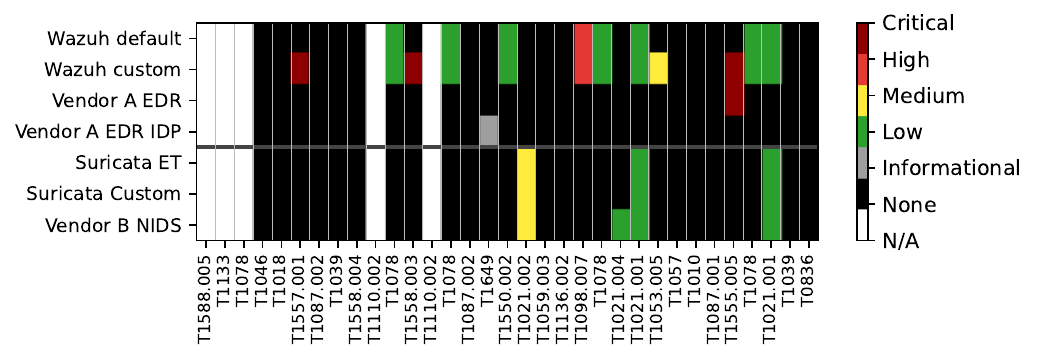}
    \label{fig:eval_team2}
  \end{subfigure}
  
  \begin{subfigure}[b]{0.485\textwidth}
    \centering
    \includegraphics[width=\linewidth]{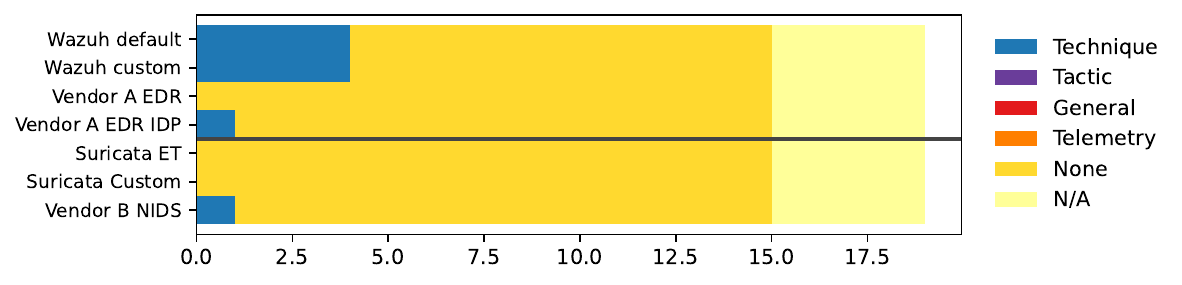}
    \label{fig:mitre_team6}
  \end{subfigure}\hfill
  \begin{subfigure}[t]{0.485\textwidth}
    \centering
    \includegraphics[width=\linewidth]{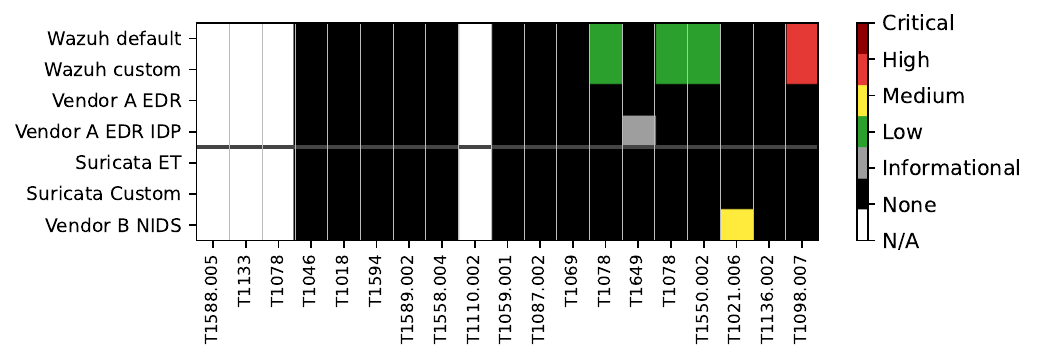}
    \label{fig:eval_team6}
  \end{subfigure}


  \begin{subfigure}[b]{0.485\textwidth}
    \centering
    \includegraphics[width=\linewidth]{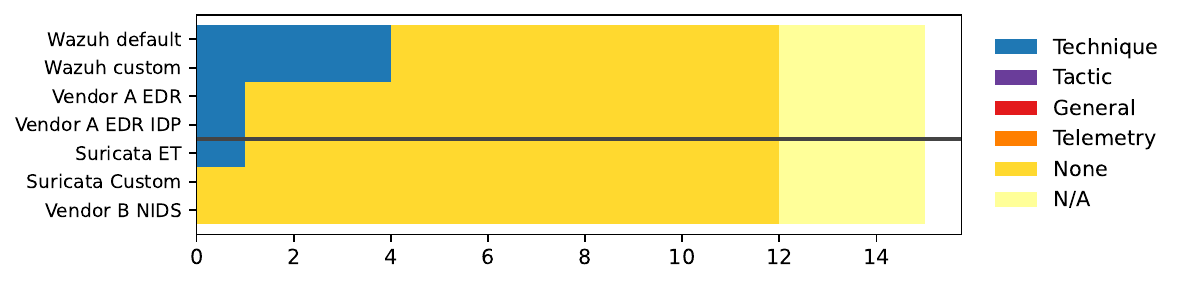} 
    \label{fig:mitre_team9}
  \end{subfigure}\hfill
  \begin{subfigure}[t]{0.485\textwidth}
    \centering
    \includegraphics[width=\linewidth]{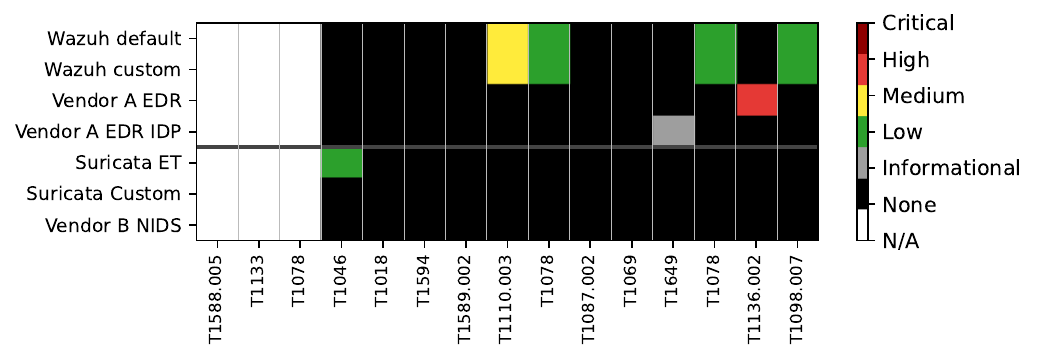}
    \label{fig:eval_team9}
  \end{subfigure}

  \caption{Comparison of MITRE Eval \cite{mitreeval} coverage (left) and alert/score profiles (right) for Teams 2 (Ent./OT Cup), 6 (Ent. Cup Winner) and 9. (Ent. Cup)}
  \label{fig:mitre_eval_grid}
\end{figure*}

Figure~\ref{fig:mitre_eval_grid} illustrates representative cases: Team~2, the only team to complete the full IT/OT kill chain; Team~6, the winner of the Enterprise Cup; and Team~9, another strong Enterprise Cup performer. 

We evaluated seven IDS configurations, covering both open-source and commercial tools:
\begin{itemize}
  \item IDS - Wazuh (default),
  \item IDS - Wazuh (custom): rules for our environment,
  \item IDS/EDR - Vendor~A~EDR\commercialmark{} (default): no additional tuning,
  \item IDS/XDR - Vendor~A~EDR\commercialmark{} (IDP): default, no additional tuning,
  \item NIDS - Suricata (ET ruleset): Emerging Threats ruleset~\cite{ETrules}, TGI ruleset~\cite{TGIrules},
  \item NIDS - Suricata (Custom): rules for our environment,
  \item NIDS - Vendor~B~NIDS\commercialmark{} (default): no additional tuning.
\end{itemize}

Beyond attack-centric analysis (attack step executed $\rightarrow$ detection observed), we also measured the false positive rate. 
Every alert from each IDS was manually labeled on a five-point confidence scale:  
(5) certainly an attack,  
(4) likely an attack,  
(3) uncertain,  
(2) likely benign,  
(1) certainly benign.  
False positives were defined as alerts labeled 1 or 2, and the false-positive rate was calculated as the fraction of all alerts falling into these categories. 
Labels were assigned manually using attacker write-ups as ground truth, and the assumption that all activity originating from the Kali foothold machine corresponded to attacker behavior. 

Although seven different configurations of IDS systems were in place, of a total of 32 techniques 11 were not alerted by any of the systems.

\begin{table}[bh]
\centering
\caption{False positive rates (FPR) per IDS configuration and team. 
Lower values indicate fewer benign events misclassified as attacks.}
\label{tab:fpr}
\begin{tabular}{lccc}
\toprule
\textbf{IDS Configuration} & \textbf{Team~2} & \textbf{Team~6} & \textbf{Team~9} \\
\midrule
Wazuh (default)       & 94.79\% & 95.30\% & 67.86\% \\
Wazuh (custom)        & 93.18\% & 92.76\% & 67.99\% \\
Vendor~A~EDR\commercialmark{} (default) & 0.00\%  & 0.00\%  & 0.00\% \\
Vendor~A~EDR\commercialmark{} (IDP)     & 0.00\%  & 0.00\%  & 0.00\% \\
Suricata (custom)     & 0.00\%  & 13.53\% & 0.00\% \\
Suricata (ET ruleset) & 0.00\%  & 2.70\%  & 0.00\% \\
Vendor~B~NIDS\commercialmark{} (default)      & 26.67\% & 33.33\% & 22.22\% \\
\bottomrule
\end{tabular}
\end{table}

This evaluation allows us to compare open-source and commercial solutions, as well as the effect of tuning versus default configurations. 
We report two primary metrics: (M1) \emph{attack-centric detection coverage}, the fraction of attack steps from writeups that triggered at least one alert; and (M2) \emph{false-positive ratio}, the fraction of alerts judged benign is shown in Table \ref{tab:fpr}. 
Together, these metrics provide a structured basis for assessing IDS performance under stealth-focused human adversaries.


\subsection{Realism of Attacker Behavior}
\label{sec:realism}
To evaluate the realism of attacker behavior in StealthCup, we compared the techniques exercised by participants against those documented in recent advanced persistent threat (APT) campaigns. 
Specifically, we used the recent CISA advisory on Volt Typhoon state-sponsored activity against U.S.\ critical infrastructure~\cite{volttyphon}, which maps observed tactics and techniques to the MITRE ATT\&CK framework. 
The comparison was conducted post-event, independent of the testbed design described in Section~\ref{sec:ill-uc}. 
We systematically assessed (i) which Volt Typhoon techniques were applicable in our IT/OT infrastructure, (ii) which were documented by participants in their attacker writeups, and (iii) which were observable in forensic artifacts (PCAPs, host logs, EDR telemetry).

The analysis reveals substantial overlap between StealthCup activity and real-world campaigns. 
Out of the Volt Typhoon techniques, 28 were directly applicable to our testbed, and all of them were exercised during the competition, as confirmed through manual forensic analysis of collected data. 
Of these, 19 techniques were described in attacker writeups and 9 were confirmed in forensic traces (e.g., Remote Service Discovery, User Discovery, Network Configuration Discovery, Process Discovery). 
Some reconnaissance techniques common in real-world intrusions (e.g., organizational or external network discovery) were not modeled in our scenario and thus not applicable in StealthCup. 
The full comparison results are provided in Appendix~\ref{app:comp_volt}.

Overall, the triangulation of applicability, participant writeups, and forensic traces demonstrates that the attacks carried out during StealthCup elicited behavior closely aligned with state-sponsored TTPs. 
This confirms that the resulting datasets and IDS evaluations are not synthetic or contrived, but grounded in realistic adversary tradecraft, making them suitable for IDS benchmarking and research.



\section{Discussion and Limitations}

\subsection{Interpretation of Results}
The evaluation results demonstrate that the majority of malicious activity remained undetected by the deployed IDS solutions across the analyzed competition runs. 
This indicates that participants were able to circumvent detection mechanisms, a behavior further encouraged by two design choices: (i) near real-time feedback on triggered detections, and (ii) the ability to reset both the infrastructure and scoring to start new runs. 
As shown in Figure~\ref{fig:solves}, steep downward changes in detection scores along the timeline suggest that teams quickly adapted their strategies once they understood the consequences of their techniques. 

The implemented attack chains in the IT domain primarily targeted Active Directory (AD). 
These attacks could be executed from an attacker machine not joined to the Windows domain and thus not fully monitored by HIDS or EDR solutions. 
This design encouraged contestants to focus on techniques detectable only by observing impact (e.g., abnormal logons, privilege escalations) rather than malware execution blocked at the endpoint. 

MITRE ATT\&CK Evaluations offer a valuable comparison point. 
While MITRE focuses on endpoint scenarios with pre-defined TTPs (e.g., the 2024 evaluations centered on ransomware and APT campaigns), StealthCup extended this approach to multi-stage AD- and OT-centric intrusions. 
By treating high-scoring contestant runs as distinct “threat actors” and their writeups as ground truth, we obtained insights into IDS performance across AD compromise and lateral movement phases that MITRE evaluations do not cover. 

False positive rates (FPR), summarized in Table~\ref{tab:fpr}, revealed striking differences between commercial and open-source systems. 
Commercial solutions such as Vendor~A~EDR\commercialmark{} and Vendor~B~NIDS\commercialmark{} produced relatively few alerts but also detected fewer attacks. 
By contrast, open-source systems (Wazuh, Suricata) generated many more alerts, with Wazuh exhibiting particularly high false-positive rates. 
This effect is partly explained by Wazuh treating benign events (e.g., successful user logons) as low-severity alerts, which inflates the alert volume and complicates analysis for SOC operators. 

\subsection{Reflections and Lessons Learned}
Because all contestants operated against identical infrastructures, they exploited the same vulnerabilities. 
Nevertheless, detection outcomes varied significantly across teams due to different tooling and attack chains. 
For instance, using \texttt{net.exe} executed via a C2 agent on a domain controller was detected by EDR, whereas executing the same command via a PsExec-like tool was not. 
Such examples illustrate both blind spots in IDS products and the brittleness of certain detection rules. 

Improving ground truth quality is critical. 
Currently, attacker write-ups were only required for the final validated run, preventing us from analyzing how techniques evolved across earlier attempts. 
Mandatory, timestamped logging of each attacker action in a structured format would enable finer-grained attribution and improve reproducibility. 

The reset-and-randomize mechanism enabled participants to explore vulnerabilities during early runs and refine their strategies in later attempts, as reflected in score progressions (Figure~\ref{fig:solves}). 
While this design lowered entry barriers and encouraged learning, it reduced the fidelity of ground truth for IDS evaluation. 
A stricter competition format with only one permitted run would provide higher-quality attack chains but substantially increase difficulty. 

Future scenarios could extend realism by modeling endpoint compromise through phishing-based initial access. 
This would require contestants to craft malware payloads delivered via phishing emails, executed on designated victim endpoints, and subsequently controlled via C2 infrastructure. 
Such an extension would more closely mirror real-world campaigns. 

Finally, the large number of low-severity alerts in Wazuh demonstrated that benign telemetry (e.g., user logons) can both aid forensic reconstruction and overwhelm SOC analysts in daily operations. 
Their utility thus depends heavily on integration with correlation rules or higher-level analytics.

Concerning the detection of individual techniques, technique T1558.004 should have been detected but it is assumed that attackers found ways to circumvent detection logic. T1039 was detected in earlier runs due to the use of honey token files, but these file seem to have been avoided by attackers in subsequent runs.

\subsection{Limitations}
Our evaluation is subject to several limitations. 

\begin{itemize}
    \item \textbf{Scenario scope.} The environment was tailored to a combined IT/OT enterprise network with AD and PLC components. While grounded in industry practice, it does not cover all attack surfaces (e.g., phishing-driven initial access, supply-chain compromise).
    \item \textbf{Ground truth completeness.} Player writeups are self-reported and may omit actions. No agent-based attacker logging was enforced, limiting traceability across all runs.
    \item \textbf{Zero-day exploitation.} StealthCup emphasized realistic misconfigurations and disclosed vulnerabilities. We did not evaluate resilience against undisclosed zero-days, which attackers in the wild may leverage.
    \item \textbf{Participant bias.} Results reflect the skill distribution and strategies of the participating penetration testers. Prize incentives sometimes shifted focus to winning objectives rather than maximizing stealth.
    \item \textbf{Randomization.} Reset and randomization did not cover all components, leaving some prerequisites of attack chains static. This may have allowed partial knowledge reuse across runs.
    \item \textbf{Evaluation breadth.} Findings stem from a single domain-specific event with limited IDS diversity (four tools, tuned vs.\ default). Broader generalization requires replication across additional infrastructures, attack chains, and detection products.
    \item \textbf{User and behavior emulation.} No real user activity was simulated, which limits evaluation of UEBA or machine-learning-based anomaly detection systems that rely on baselines of “normal” behavior.
\end{itemize}

Despite these limitations, StealthCup demonstrated that realistic, multi-team attack scenarios can expose IDS blind spots not visible in static benchmarks. 
We further observed that providing a realistic and modern environment motivated experienced penetration testers to participate—often in mixed teams where senior players collaborated with less experienced ones—supporting both knowledge transfer and the exploration of creative evasion strategies.


\section{Conclusion}
This first StealthCup demonstrated that an evasion-focused CTF can generate realistic attack traces, and comparative insights into IDS performance. 
By combining reproducible IT/OT infrastructures with professional penetration testers and a stealth-oriented scoring system, we showed that StealthCup elicits attacker behavior closely aligned with documented APT campaigns while revealing blind spots across both open-source and commercial IDS solutions. 

Future iterations will expand StealthCup along several dimensions. 
First, stronger ground truth will be ensured by introducing optional agents to log attacker activity in detail—balancing accuracy with participant acceptance. 
Second, involving blue teams in scoring and analysis can better approximate operational settings. 
Third, expanding the number of initial intrusion vectors, objectives, and IDS configurations will increase coverage and strengthen generality. 
Together, these steps will advance StealthCup from a domain-specific case study toward a reusable methodology for benchmarking IDS under stealth-focused, human-driven attacks.

\bibliographystyle{ACM-Reference-Format}
\bibliography{bib}

\appendix
\section{Comparison of Volt Typhon Techniques with Infrastructure and Writeups}
\label{app:comp_volt}
\begin{table*}[t]
\centering
\scriptsize
\caption{Volt Typhoon techniques vs.\ StealthCup: applicability (App), forensic evidence (For), and count of team writeups mentioning the technique (Wr).}
\label{tab:volt-mapping}
\begin{tabular}{l p{7.5cm} c c c c}
\toprule
\textbf{Tactic} & \textbf{Technique} & \textbf{VT} & \textbf{App} & \textbf{For} & \textbf{Wr} \\
\midrule
Reconaissance & Search Victim-Owned Websites - T1594 & \checkmark & \checkmark & \checkmark & 10 \\
 & Gather Victim Identity Information: Email Addresses - T1589.002 & \checkmark & \checkmark & \checkmark & 10 \\
 & Gather Victim Org Information - T1591 & \checkmark &  &  & 0 \\
 & Gather Victim Network Information - T1590 & \checkmark &  &  & 0 \\
 & Gather Victim Identity Information - T1589 & \checkmark & \checkmark & \checkmark & 10 \\
 & Search Open Websites/Domains - T1593 & \checkmark & \checkmark & \checkmark & 10 \\
 & Gather Victim Host Information - T1592 & \checkmark &  &  & 0 \\
Resource Development & Obtain Capabilities: Exploits - T1588.005 & \checkmark & \checkmark & \checkmark & 13 \\
 & Acquire Infrastructure: Botnet - T1583.005 & \checkmark &  &  & 0 \\
 & Compromise Infrastructure: Botnet - T1584.005 & \checkmark &  &  & 0 \\
 & Compromise Infrastructure: Server - T1584.004 & \checkmark &  &  & 0 \\
Initial Access & External Remote Services T1133 & \checkmark & \checkmark & \checkmark & 13 \\
 & Valid Accounts T1078 & \checkmark & \checkmark & \checkmark & 14 \\
 & Exploit Public-Facing Application T1190 & \checkmark &  &  & 0 \\
Execution & Command and Scripting Interpreter PowerShell - T1059.00  & \checkmark & \checkmark & \checkmark & 1 \\
 & Command and Scripting Interpreter: Windows Command Shell - T1059.003 &  & \checkmark & \checkmark & 6 \\
 & Windows Management Instrumentation - T1047 & \checkmark & \checkmark & \checkmark & 0 \\
 & Command and Scripting Interpreter: Unix Shell T1059.004 & \checkmark & \checkmark & \checkmark & 0 \\
Persistence & Modify Registry - T1112 & \checkmark & \checkmark & \checkmark & 0 \\
 & Create Account: Domain Account - T1136.002 &  & \checkmark & \checkmark & 10 \\
 & Account Manipulation: Additional Local or Domain Groups - T1098.007 &  & \checkmark & \checkmark & 11 \\
Privilege Escalation & Exploitation for Privilege Escalation - T106  & \checkmark & \checkmark & \checkmark & 2 \\
 & Scheduled Task/Job: Scheduled Task - T1053.005 &  & \checkmark & \checkmark & 3 \\
Credential Access & Brute Force: Password Cracking - T1110.002 & \checkmark & \checkmark & \checkmark & 5 \\
 & Steal or Forge Kerberos Tickets: AS-REP Roasting - T1558.004 &  & \checkmark & \checkmark & 4 \\
 & Steal or Forge Kerberos Tickets: Kerberoasting - T1558.003 &  & \checkmark & \checkmark & 2 \\
 & OS Credential Dumping: NTDS T1003.003 & \checkmark & \checkmark & \checkmark & 2 \\
 & Unsecured Credentials T1552 & \checkmark & \checkmark & \checkmark & 1 \\
 & Credentials from Password Stores: Credentials from Web Browsers - T1555.003 & \checkmark & \checkmark & \checkmark & 0 \\
 & Adversary-in-the-Middle: LLMNR/NBT-NS Poisoning and SMB Relay - T1557.001 &  & \checkmark & \checkmark & 4 \\
 & Steal or Forge Authentication Certificates - T1649 &  & \checkmark & \checkmark & 10 \\
 & Credentials from Password Stores: Password Managers - T1555.005 &  & \checkmark & \checkmark & 1 \\
Discovery & Network Service Discovery - T1046 & \checkmark & \checkmark & \checkmark & 11 \\
 & Remote System Discovery - T1018 &  & \checkmark & \checkmark & 0 \\
 & Account Discovery: Domain Account - T1087.002 & \checkmark & \checkmark & \checkmark & 8 \\
 & Permission Groups Discovery - T1069 & \checkmark & \checkmark & \checkmark & 6 \\
 & Connection Proxy - T1090.002 & \checkmark &  &  & 0 \\
 & System Owner/User Discovery - T1033 & \checkmark & \checkmark & \checkmark & 0 \\
 & System Network Configuration Discovery - T1016 & \checkmark & \checkmark & \checkmark & 0 \\
 & Indicator Removal: File Deletion - T1070.004 & \checkmark & \checkmark & \checkmark & 0 \\
 & Application Window Discovery - T1010 & \checkmark &  &  & 0 \\
 & System Service Discovery - T1007 & \checkmark & \checkmark & \checkmark & 1 \\
 & File and Directory Discovery - T1083 & \checkmark & \checkmark & \checkmark & 5 \\
 & Process Discovery - T1057 &  & \checkmark & \checkmark & 1 \\
 & Account Discovery: Local Account - T1087.001 & \checkmark & \checkmark & \checkmark & 0 \\
Lateral Movement & Use Alternate Authentication Material: Pass-the-Hash - T1550.002 & \checkmark & \checkmark & \checkmark & 9 \\
 & Remote Services: SMB/Windows Admin Shares - T1021.002 &  & \checkmark & \checkmark & 3 \\
 & Remote Services: Remote Desktop Protocol - T1021.001 & \checkmark & \checkmark & \checkmark & 3 \\
 & Use Alternate Authentication Material: Pass the Ticket - T1550.003 & \checkmark & \checkmark & \checkmark & 0 \\
 & Remote Service Session Hijacking - T1563 & \checkmark &  &  & 0 \\
 & Remote Services: Cloud Services T1021.007 & \checkmark &  &  & 0 \\
 & Remote Services: SSH - T1021.004 &  & \checkmark & \checkmark & 14 \\
 & Remote Services: Windows Remote Management - T1021.006 &  & \checkmark & \checkmark & 2 \\
\bottomrule
\end{tabular}
\end{table*}

\end{document}

%% file: defs.tex
\usepackage{xparse}
\newcommand{\bnm}{\begin{newmath}}
\newcommand{\enm}{\end{newmath}}

\newcommand{\bea}{\begin{eqnarray*}}%
\newcommand{\eea}{\end{eqnarray*}}%

\newcommand{\bne}{\begin{newequation}}
\newcommand{\ene}{\end{newequation}}

\newcommand{\bal}{\begin{newalign}}
\newcommand{\eal}{\end{newalign}}

\newenvironment{newalign}{\begin{align}%
\setlength{\abovedisplayskip}{4pt}%
\setlength{\belowdisplayskip}{4pt}%
\setlength{\abovedisplayshortskip}{6pt}%
\setlength{\belowdisplayshortskip}{6pt} }{\end{align}}

\newenvironment{newmath}{\begin{displaymath}%
\setlength{\abovedisplayskip}{4pt}%
\setlength{\belowdisplayskip}{4pt}%
\setlength{\abovedisplayshortskip}{6pt}%
\setlength{\belowdisplayshortskip}{6pt} }{\end{displaymath}}

\newenvironment{newequation}{\begin{equation}%
\setlength{\abovedisplayskip}{4pt}%
\setlength{\belowdisplayskip}{4pt}%
\setlength{\abovedisplayshortskip}{6pt}%
\setlength{\belowdisplayshortskip}{6pt} }{\end{equation}}

\newcounter{ctr}

\newcounter{mytable}
\def\mytable{\begin{centering}\refstepcounter{mytable}}
\def\endmytable{\end{centering}}

\newcounter{myfig}
\def\myfig{\begin{centering}\refstepcounter{myfig}}
\def\endmyfig{\end{centering}}

\newlength{\saveparindent}
\newlength{\saveparskip}
\newcommand{\E}{{\rm I\kern-.3em E}}

\renewcommand{\eqref}[1]{\mbox{Equation~(\ref{#1})}}

\def \part {part}

\renewcommand{\paragraph}[1]{\vspace*{6pt}\noindent\textbf{#1}\;}

\def \blackslug{\hbox{\hskip 1pt \vrule width 4pt height 8pt
    depth 1.5pt \hskip 1pt}}
\def \qed{\quad\blackslug\lower 8.5pt\null\par}

\newcounter{mynote}[section]

\newcommand\ignore[1]{}

\newcounter{rcnote}[section]

\newcounter{mrnote}[section]

\newcounter{fknote}[section]

\newcounter{anote}[section]

\DeclareMathSymbol{\mlq}{\mathord}{operators}{``}
\DeclareMathSymbol{\mrq}{\mathord}{operators}{`'}

\newcommand{\rhf}[2]{R_{f, \gamma}}

\DeclareDocumentCommand{\edist}{o o}{
  \ensuremath{
    \IfNoValueTF{#1}{{d}}{{\sf d}(#1,#2)}
  }
}

\newcommand{\olrk}[1]{\ifx\nursymbol#1\else\!\!\mskip4.5mu plus 0.5mu\left(\mskip0.5mu plus0.5mu #1\mskip1.5mu plus0.5mu \right)\fi}

\NewDocumentCommand{\indseq}{ O{1} O{r} }{{#1}\ldots {#2}}
